# STRUCTURAL HEALTH MONITORING WITH DISTRIBUTED WIRELESS SENSOR NETWORKS

**Daniel-Ioan Curiac, Ovidiu Banias, Ioan Borza**
"Politehnica" University of Timisoara, Romania

Abstract – *Wireless Sensor Networks(WSN) are a today technology with great practicability in the real world. We focus on describing WSN architecture, regarding usefulness in constructions like structural health monitoring and importance, and advantages of using WSN in this domain.*

*Rezumat – Retelele de senzori reprezinta o tehnologie a zilelor noastre cu mare aplicabilitate in domenii reale. In acest articol ne indreptam atentia asupra arhitecturii retelelor de senzori privind importanta lor in detectarea si monitorizarea rezistentei structurilor.Vor fi prezentate importanta si avantajele utilizarii retelelor de senzori in domeniul constructiilor cat si exemple sugestive.*

## I. INTRODUCTION

A sensor network is a network of small distributed devices that are using sensors for measurement (temperature, motion, pressure, sound) and for prediction (weather forecast, fire ignition, earthquakes, military attack, building safety). These small devices are called motes and are composed from board, sensors, radio transmitter/receptor and power supply (batteries in most cases), some of them being equipped with a small processor for data processing and memory for storage. The motes are deployed in different kind of environments being able to self organize themselves in a hierarchical sensor network and produce the gathered information to the receiver. The main advantage is the small size of the motes (the size of a coin) combined with the small cost per piece and the capacity to self organize not depending of a previous setup and configuration.

Even if sensor networks research started in 1980 at the Defense Advanced Research Project Agency (DARPA) with the Distributed Sensor Networks project, beginning with year 2000 we can say this kind of networks began to be actively researched [1]. The last years were very important in sensor networks research because of the rapid growth of technology and increasing commercial interest in this kind of solutions. The growth of technology brought smaller motes (smaller boards, processors and sensors) and longer battery life, but even so, optimization of the energy consumption remains one of the major problems of interest. Being deployed in all kind of environments the mote's life depends on the battery, this is why the more optimized the power consumption algorithms, the more life time of the mote. Also, for the sensor network to worth the money regarding the quite big number of motes needed (100 or maybe 1000), these should have small prices, to afford loses of motes in time because of the battery usage and of course to afford deploying new ones without becoming to expensive.

Some of the motes are simple devices equipped only with sensors and radio transmitters, used just for gathering and transmitting information. Other motes are more complex, equipped with processors and memory being meant for receiving information for simple motes, aggregate it and than pass it forward to other receiving motes until the information arrives to the base (server).

## II. WIRELESS SENSOR NETWORKS MOTE ARCHITECTURE.

The mote architecture [2] is presented in Fig. 1: power supply unit, sensing unit, processing unit and radio transceiver unit. Power supply unit is composed mostly from disposable batteries or scavenging solar cells. The sensing unit receives the information from the environment through the sensors and then with the ADC (Analog to Digital Converter) transformed sensed information in a digital format to pass forward to the processing unit. The processing unit is the brain of the mote, being composed from a micro processor running on a real time operating system, memory and algorithms. Sensed information is analyzed and processed, possibly stored in the memory, and then according to established protocols, sent through the radio transmitter to neighbor motes.

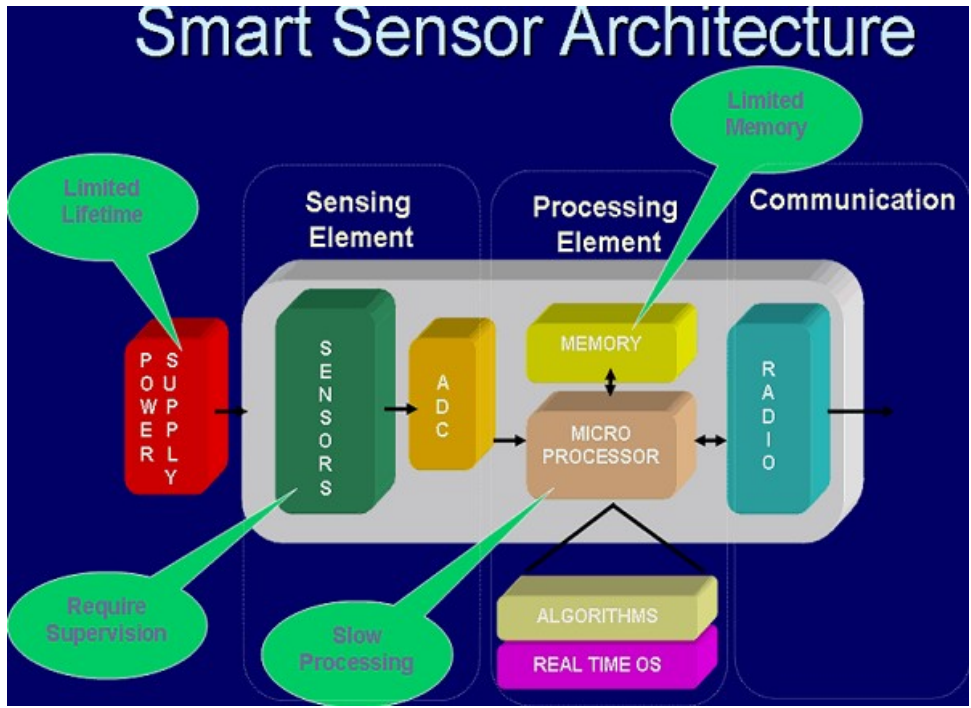

Fig. 1: Smart Sensor Architecture

Nowadays MICA2 (Table 1) mote of Crossbow and Berkeley Labs is one of the most appreciated mote on the market being updated continuously. The main advantage of MICA 2 is that is an open hardware and software platform for smart sensing, being able to plug-in sensors boards. The main advantage of this mote is that it could be used in almost any environment with little power consumption and a battery life of more than 1 year. The radio range could be modified depending on certain deployment needs for less power consumption and more battery life. In wireless sensor networks used for detection (fire, flood, object movement, etc.) the information validity should be no more than few seconds, enough to be propagated to the sink (the sink of a wireless sensor network is a device that reports gathered information to monitoring groups or fire an alarm). In this situation, communication protocol security is not so important and this could be used for less computation, so less power consumption.

| Components | Characteristics | Remarks |
|---|---|---|
| CPU | ATmega128L, 7.4 MHz | |
| Data MEM | 512 KB | |
| Program MEM | 128 KB | |
| AD converter | 10 bit | |
| Processor current draw | 8 mA | active mode |
| | <15 µA | sleep mode |
| Radio frequency | 315/433/868/916 MHz | |
| Data Rate | 38.4 baud | |
| Radio range | 3 Km | |
| Radio current draw | 25 mA | transmit |
| | 8 mA | receive |
| | <1 µA | sleep |
| Power | 2 AA batteries | |
| External power | 2.7 – 3.3 V | |
| Size | 58 x 32 x 7 mm | |
| Weight | 18 g | |

Table 1

## III. STRUCTURE MONITORING

Buildings are exposed to natural hazards (fire ignition, earthquakes, strong winds, construction material problems) and wireless sensor networks could be a cheap but very useful technology for detection and prevention. The old technology of wired data acquisition systems for collecting structural vibrations is far less reliable and more expensive [4]. Because of their small size, the motes could be easily deployed in the walls for example in a certain capsule. The only problem in this case could be the mote's life time if there will be no possibility to reach it and exchange the battery, but in many cases the mote being deployed within a capsule with a known place, the capsule could be removed and the battery exchanged.

Bridges also can be subject to natural hazards and detecting structural problems is a subject of maximum interest. More and more bridges today are equipped with wireless monitoring systems [3], every minor change in material property or bridge problems being detected and reported wireless (Fig. 2).

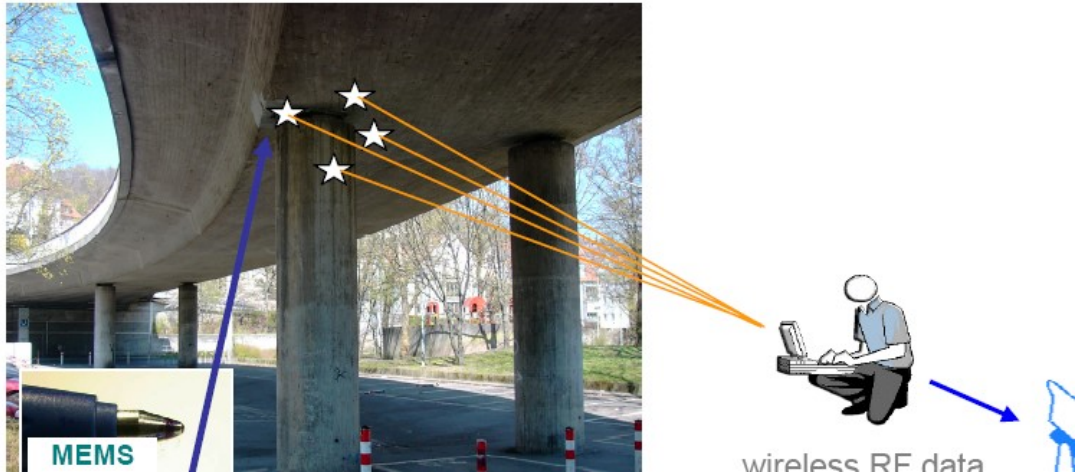
Fig. 2. Wireless Sensor Network for bridge measurements

The main advantage of the motes is that they are very small, can report immediately any problem sensed through the incorporated sensors and a potential problem can be announced at the control center in couple of seconds. Detection is essential and is the first step to repairing. The sooner the problem is detected the bigger the chance to repair it.

The data aggregation is very important for energy saving, for example a complex mote equipped with processor and memory should first aggregate temperature information received from sensors in his area, making an average of temperature in last hour and then send it forward besides sending each temperature received from the sensing motes every second or minute. Big amount of energy is saved this way and the network life is increased. For motes to be able to aggregate data, specific algorithms and artificial intelligence are used and implemented in order to achieve low power consumption regarding to the amount of used computational force. Without aggregation each sensing mote will send information to all his neighbors and receive information from the neighbors and pass it forward, this way one piece of information will be transmitted and retransmitted a lot more times than using the aggregation technique, and a lot of energy will be wasted [5].

Enormous damage produced by natural disasters like earthquakes, fire, flood, could be avoided or at least detected in short time, minimizing as low as possible the human and material losses. For this kind of problems the sensor networks are an affordable technology not hard to implement and deploy in needed environments [6].

The sensor networks motes should be able to self organize themselves. The capacity of self organization is essential because the motes need to work in almost any environment that can't offer the possibility to route information through the motes like in Local Area Networks for example, the routing technique couldn't be designed by a network administrator. If one mote's battery become empty than the network should be able to remove that mote from the network topology and if one mote is deployed in an already organized sensor network, this one should be able to communicate with the network and to be accepted and entered in the network topology.

In the literature there are solutions based on small video cameras [7], witch send frames of the area generating the problems after earlier sensor detection. This kind of motes are useful is certain situations when the problem need to be visually located. In most cases the motes of the sensor network are deployed into the walls or attached to the structure. It is necessary to deploy more than one mote in a certain area, because in case of damage of empty battery, another mote to be able to detect and report possible problems.

## IV. CONCLUSIONS

The application of these networks is very wide and still open, being a useful technology for all kind of domains including seismic detection, fire ignition, habitat monitoring, structural health monitoring, home and office applications. In the next years will be developed smaller motes with less power consumption implying greater battery life. More and more buildings and bridges will be equipped with sensor networks technology for detecting structural health problems, reducing costs and increasing benefits.